\documentclass[%
 reprint,
 amsmath,amssymb,
 aps,
]{revtex4-1}

\usepackage{graphicx}
\usepackage{dcolumn}
\usepackage{bm}

\begin{document}

\preprint{APS/123-QED}

\title{Constraining the general reheating phase in the $\alpha$-attractor inflationary cosmology}

\author{Alessandro Di Marco}
\email{alessandro.di.marco@roma2.infn.it}
\author{Paolo Cabella}
\author{Nicola Vittorio}
\affiliation{%
University of Rome - Tor Vergata, Via della Ricerca Scientifica 1\\
INFN Sezione di Roma “Tor Vergata”, via della Ricerca Scientifica 1, 00133 Roma, Italy}%


\begin{abstract}

In this paper we constrain some aspects of the general postinflationary phase in the context of superconformal $\alpha$-attractor models of inflation.
In particular, we provide constraints on the duration of the reheating process, $N_{reh}$, and on the reheating temperature, $T_{reh}$,
simulating possible and future results given by the next-generation of cosmological missions.
Moreover, we stress what kinds of equation-of-state parameter, $w_{reh}$, are favored for different scenarios.
The analysis does not depend on the details of the reheating phase and it is performed assuming different measurements of the tensor-to-scalar ratio $r$.

\end{abstract}

\pacs{Valid PACS appear here}
\maketitle


\section{Introduction}

The theory of inflation is the most promising process for the description of the early Universe \cite{1}.
The detection of the inflationary background of gravitational waves (GW) could provide a further confirmation of the goodness of the inflationary paradigm
and represents one of the most important goals of the current research activity.
The Planck mission constraints $r<0.07$ at $95\%$ C.L. on a pivot scale of $k_*=0.002$ Mpc$^{-1}$ \cite{2,3}.
In the next future, we might detect inflationary gravitational waves by forthcoming polarization missions of the Cosmic Macrowave Background (CMB) 
(like LITEBIRD or COrE, see \cite{4,5,6,7,8} for details) or by gravitational waves experiments (ALIA and BBO missions, see \cite{9} for a review).
However, the knowledge of the inflationary process could shed light on several aspects of the postinflationary phase, the reheating phase, responsible for
the production of the observable entropy in the Universe
\cite{10,11,12,13,14,15,16,17,18,19,20,21}.
In this work, we study what kinds of constraints we may have on some aspects of the postaccelerated phase
assuming future detections of the tensor-to-scalar ratio $r$ in the range of $10^{-2}$-$10^{-3}$.
In particular, we assume the $\alpha$-attractor models of inflation (see \cite{22,23,24,25,26,27} for details) as the model for the early Universe and we provide
constraints on the macro reheating variables, i.e., the number of $e$-foldings during the reheating phase, $N_{reh}$, and the reheating temperature, $T_{reh}$.
In doing this, we outline the possible mean values of the equation-of-state parameter (EoSp), $w_{reh}$, compatible with each given observation.
Furthermore, we show how the constraints on these parameters are sensible to the uncertainty on the scalar spectral index, $n_s$.
Currently, PLANCK mission fixes $\sigma_{n_s}=0.006$ (see \cite{2,3}) which provides a very large uncertainty on $T_{reh}$.
The paper is organized in the following way.
In Sec. II, we recall the basic concepts of inflation and discuss the reheating phase introducing the fundamental parameters, $N_{reh}$, $T_{reh}$ and $w_{reh}$.
In Sec. III, we reconstruct the main parameters of the reheating stage with respect to the mean value of the EoSp of the reheating fluid.
In Sec. IV, we discuss the results and the implications of the work.
In this paper we use the natural units of particle and cosmology $\hbar=c=k_B=1$.

\section{Inflation, Reheating and Potential reconstruction}

\subsubsection{Inflation and quantum fluctuations}

The inflationary paradigm provides a fast accelerated expansion of the early Universe, on energy scale close to the one of the simmetry breaking of the grand unified
interaction ($E_{inf} < 10^{16}$ GeV).
Mathematically, we have $a(t)\sim e^N$, where $a(t)$ is the cosmic scale factor and $N$ is the number of $e$-foldings, the number of exponential growth before the end of inflation.
The most popular scenario for inflation involves the simplest version of a scalar field (neutral, homogeneous, minimally coupled to gravity and
canonically normalized) with an effective autointeraction potential sufficiently flat and characterized by a global minimum.
Then, the early-cosmological action assumes the following generic form:
\begin{eqnarray}
S=\int d^4x \sqrt{-g} \left\{\frac{1}{2}M^2_p R - \frac{1}{2}g_{\mu\nu}\partial^{\mu}\phi \partial^{\nu}\phi -V(\phi)   \right\}
\end{eqnarray}
where $R$ is the Ricci scalar, $g_{\mu\nu}$ is the metric tensor, $g$ is its determinant and $M_p$ is the reduced Planck mass.
The scalar field, $\phi$ is often called ``inflaton'', while the function $V(\phi)$ is the inflationary potential.
The nature of the inflaton field is an open issue.
However, it is common belief that $\phi$ could arise from the low energy limit of some more fundamental cosmological theory.
The inflationary process starts with the scalar field that explores the flat ragion or ``false vacuum" 
with a very slow dynamics: $\partial_{\mu}\phi\partial^{\mu}\phi \ll V(\phi)$.
Useful quantities to describe this phase are the so called slow roll-parameters,
\begin{eqnarray}\label{eqn: psrp }
\epsilon_V(\phi)=\frac{M^2_p}{2} \left(\frac{V'(\phi)}{V(\phi)} \right)^2,  \quad \eta_V(\phi)= M^2_p \left( \frac{V''(\phi)}{V(\phi)} \right)
\end{eqnarray}
During this stage we have the production of the cosmological perturbations related to the large scale structure and to the temperature fluctuations
of the cosmic background radiation \cite{28,29,30,31,32,33,34,35,36}.
The power spectrum and the spectral index for the generated scalar perturbations are respectively,
\begin{eqnarray}\label{eqn: osservabili1}
P_s(k)= \frac{1}{8\pi^2  M_p^2}\frac{H^2}{\epsilon_V}\Bigl|_{k_*}  
\end{eqnarray}
and
\begin{eqnarray}
n_s=1-6\epsilon_V+2\eta_V
\end{eqnarray}
while the corresponding quantities for the tensor sector are
\begin{eqnarray}\label{eqn: osservabili2}
P_t(k)=\frac{2}{\pi^2}\frac{H^2}{M^2_p}\Bigl|_{k_*}
\end{eqnarray}
and
\begin{eqnarray}
n_t=-2\epsilon_V
\end{eqnarray}
Finally, the tensor-to-scalar ratio amplitude is given by
\begin{eqnarray}\label{eqn: ratio}
r=\frac{P_s(k)}{P_t(k)}=16\epsilon_V
\end{eqnarray}
Note that all of these quantities are functions of the scalar field and are calculated at first order in the slow roll parameters at horizon crossing of the pivot scale $k_*$.
The observables $n_s$ and $r$ are the quantities useful to test the ``degrees of compatibility'' of a given model $V(\phi)$ with the cosmological data.
As broadly discussed in \cite{37,38}, we can rewrite the observables in terms of the number of $e$-foldings $N_*$, by using the relation
\begin{eqnarray}\label{eqn: eqmotion}
N(\phi_*,\phi_{end},\beta_i)=\frac{1}{M_p}\int_{\Delta\phi} d\phi \frac{1}{\sqrt{2\epsilon_V(\phi)}}
\end{eqnarray} 
where $\Delta\phi=\phi_* - \phi_{end}$ is the variation of the field between the horizon crossing of quantum mode $k_*$ and the end of inflation, while 
$\beta_i$ are the parameters of the potential function.
In conclusion, we have relations of the form $n_s=n_s(N_*)$, $r=r(N_*)$.\\

\subsubsection{The end of inflation: the reheating phase}

The inflationary process stops when the field approaches the value $\phi_{end}$.
In general, one has $\epsilon_V\sim 1$.
It is straightforward to show that the energy density at this epoch is given by
\begin{eqnarray}\label{eqn: energydensity}
\rho(\phi_{end})\simeq\frac{4}{3}V(\phi_{end})
\end{eqnarray}
In the following, the field quickly relaxes toward the real minimum of the potential and here it takes mass, as well as starts to oscillate and decay, producing the radiation
characterizing the radiation dominated era of standard Big Bang Model.
This phase is called the reheating phase and can be described by a single reheating fluid, with a ``mean'' EoSp, $w_{reh}$.
The physics of reheating is extremely complicated and depends on many factors such as the dominant operator terms of the potential about the minimum
and the details of the inflaton physics.
The first and simplest proposed model for the description of the reheating era is sometimes called ``elementary theory of reheating'' \cite{9,10,11}
although we can have more complicated effects like preheating and turbulent stages \cite{13,14,15,16,17,18,19}
(a complete review can be found in \cite{20} while the first reheating constraints are given in \cite{21}).
The simplest model provides a first stage of coherent oscillations of the inflaton field about the minimum of a simple quadratic potential.
The value of EoSp is fixed to be zero, $w_{reh}=0$.
The oscillations produce a cold gas of inflaton-particles that decay to relativistic particles.
Afterwards, a thermalization stage takes place in which the particles strongly interact with each other and reach the thermal equilibrium. 
The duration $\tau_{reh}$ of the reheating phase is of the order of  $\tau_{reh}\sim \Gamma_{\phi}^{-1}$, where $\Gamma_{\phi}$
is the inflaton decay rate. 
However, independent on the details of the reheating mechanism we can introduce two macro quantities: the number of $e$-foldings during the reheating stage, $N_{reh}$
and the temperature at the end of the process, $T_{reh}$.
Indeed, exploiting the fact that the number of $e$-foldings before the end of inflation depends 
on the complete history of the Universe after the inflationary phase (and so on $N_{reh}$), 
we can show that for any model $V(\phi)$ holds (see references \cite{39,40,41,42}),
\begin{eqnarray}\label{eqn: reh}
N_{reh}=\frac{4}{1-3 w_{reh}}f(\beta_i,O_i,N_*)
\end{eqnarray}
Here, $\beta_i$ are the parameters of the model and $O_i$ are the known cosmological quantities (CMB photon temperature $T_0$, Hubble rate $H_0$,...).
The function $f$ is defined as
\begin{align}\label{eqn: func f}
f(\beta_i,O_i,N_*)&=\left[ -N_* - \ln\left( \frac{k_*}{a_0 H_0} \right) + \ln\left( \frac{T_0}{H_0} \right) \right] \nonumber\\
&+\left[ \frac{1}{4}\ln\left( \frac{V^2_*}{M^4_p \rho_{end}} \right) -\frac{1}{12}\ln(g_{reh})  \right] \nonumber\\
&+\left[ \frac{1}{4}\ln\left(\frac{1}{9}\right) + \ln\left( \frac{43}{11}\right)^{\frac{1}{3}} \left(\frac{\pi^2}{30} \right)^{\frac{1}{4}}\right] .
\end{align}
where $w_{reh}$ is the effective EoSp of the conserved ``reheating fluid", $k_*$ is the pivot scale ($k_*=0.002$ Mpc$^{-1}$, in our case), $N_*$ is
the remaining number of $e$-foldings before the end of inflation when $k_*$ crosses the horizon, $H_0=1.75\times 10^{-42}$ GeV is the current Hubble rate,
$T_0=2.3\times 10^{-12}$ GeV is the CMB temperature, $V_*$ is the energy density when $k_*$ leaves the horizon, $\rho_{end}$ is the energy density at the end of inflation and $g_{reh}\sim 100$ is the number of relativistic degrees of freedom at the onset of the radiation dominance.
Now, the above equations allow to derive the duration of the reheating phase, once a model is chosen.
Note that, when the function $f$ equals zero, the reheating occurs instantaneously providing the maximum value for $N_*$.
Eq.($\ref{eqn: reh}$) tells us that the duration of the reheating era depends on the mean value of the EoSp, i.e, on the nature of the reheating fluid.
In particular, we should note that if $f>0$ then $N_{reh}>0$ if $w_{reh}<1/3$.
Alternatively, if $f<0$ we need $w_{reh}>1/3$.
Actually, we need for $w_{reh}>-1/3$ to exit from the inflationary phase.
On the other side, we need $w_{reh}<1/3$ to avoid $\sim \phi^{6}$ close to the minimum of $V(\phi)$.
Then, this second case appears unnatural in the context of quantum field theory although remain a possibility \cite{39}.
In literature one can find different numerical studies on this question, such as \cite{21}.
The parameter $N_{reh}$ allows to calculate the ``general reheating temperature" (GRT) $T_{GR}$ that results in
\begin{eqnarray}\label{eqn: gen_temp}
T_{GR}=\left( \frac{40 V_{end}}{\pi^2 g_{reh}} \right)^{1/4} \exp{ \left[-\frac{3}{4}(1+w_{eff})N_{reh}\right]}
\end{eqnarray}
In the limit of $N_{reh}\rightarrow 0$, the instantaneous scenario for the reheating mechanism is recovered.
In this case, the ``instantaneous reheating temperature" (IRT) $T_{IR}$ comes out to be
\begin{eqnarray}\label{eqn: inst_temp}
T_{IR}=\left( \frac{40 V_{end}}{\pi^2 g_{reh}} \right)^{1/4}
\end{eqnarray}
Then, Eq.($\ref{eqn: gen_temp}$) shows an exponential modulation of the instantaneous case.
As we can expect, the energy scale of the reheating gets lower as its duration becomes larger.
Moreover, the final reheating temperature also depends on $w_{reh}$.
Now, the inflationary observables depend on the $N_*$ and on the value of the EoSp by the inverting Eq.($\ref{eqn: reh}$).
In this sense (once a model $V(\phi)$ is chosen and for fixed values of the cosmological parameters in Eq.($\ref{eqn: reh}$)), different values of the EoSp provide different pairs $(n_s,r)$.
On the other hand, measurements of both $n_s$ and $r$ allow to reconstruct a given model $V(\phi)$ (see \cite{37} for details), and from it $N_{reh}$ and $T_{reh}$
using the equations Eq.($\ref{eqn: reh}$) and Eq.($\ref{eqn: gen_temp}$).
In the following, we apply such a procedure in the context of the $\alpha$-attractor inflationary models.

\section{The $\alpha$-attractor models of inflation: an overview}

\begin{figure}[htbp]
\centering
\includegraphics[width=8.6cm, height=6cm]{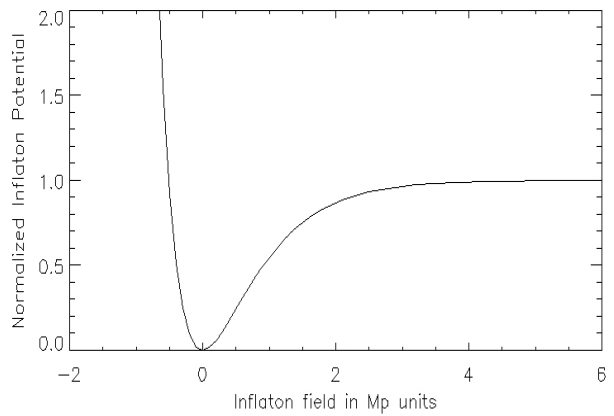}
\caption{Typical shape of an $\textbf{E-model}$ $\alpha$-attractor function.}
\label{fig: shape}
\end{figure}

The $\alpha$-attractor models of inflation represent a very interesting class of inflationary models.
This class can rise from different scenarios, but the most advanced version is actually realized in the supergravity landscape
(\cite{22,23,24,25,26,27}).
In particular we have two classes of $\alpha$-attractor potentials.
The first one is the so-called T-model class for which the prototype potential
is given by
\begin{eqnarray}
V(\phi)=\Lambda^4 \tanh^{2n}\left(\frac{\phi}{\sqrt{6\alpha} M_p}\right)
\end{eqnarray}
The second is the E-model class defined by
\begin{eqnarray}
V(\phi)=\Lambda^4 \left( 1 - e^{b\phi/M_p} \right)^{2n}
\end{eqnarray}
These classes of models are particular interesting especially by virtue of its capability to interpolate a broad range of predictions in the $(n_s,r)$-plane. 
Herein, we want to focus on the simplest version of the E-class, where $n=1$
\begin{eqnarray}
V(\phi)=\Lambda^4\left(1-e^{-b \phi/M_p}\right)^2
\end{eqnarray}
In general, $b$ is the model parameter defined as:
\begin{eqnarray}
b=\sqrt{\frac{2}{3\alpha}}
\end{eqnarray}
In such a way, within supergravity context, the $\alpha$ parameter is related to the curvature of the K{\"a}lher geometry associated with the inflaton field
\begin{eqnarray}\label{eqn: kaler}
R_K=-\frac{2}{3\alpha}
\end{eqnarray}
In the limit of $b\phi/M_p>>1$ the false vacuum is recovered as shown in Fig.($\ref{fig: shape}$).

The relation between the vacuum $\phi_*$ and the number of $e$-foldings $N_*$ is given by solving the equation of motion Eq.($\ref{eqn: eqmotion}$):
\begin{eqnarray}\label{eqn: efolds complete}
N_*=\frac{1}{2b^2}\left( e^{b\frac{\phi_*}{M_p}} -  e^{b\frac{\phi_{end}}{M_p}}   \right)-\frac{1}{2b}\left(   \frac{\phi_*}{M_p}-\frac{\phi_{end}}{M_p} \right).
\end{eqnarray}
Note that, when we apply the condition $b\phi_*/M_p>>1$, we can simplify the solution as
\begin{eqnarray}\label{eqn: efolds app}
N_*\simeq \frac{1}{2b^2}e^{b\frac{\phi_*}{M_p}}.
\end{eqnarray}
Using the condition $\epsilon_V(\phi_{end})\sim 1$, we can show the vacuum expectation value of the scalar field
at the end of the accelerated phase, results in
\begin{eqnarray}\label{eqn: vev end}
\frac{\phi_{end}}{M_p}=\frac{1}{b}\ln\left(1+\sqrt{2}b\right).
\end{eqnarray}
This value is crucial to determine $\phi_*$ once $N_*$ is fixed, inverting Eq.($\ref{eqn: efolds complete}$) by the Lambert function.
The predicted scalar spectral index and the tensor-to-scalar ratio of the $\alpha$-attractor models (in the limit of small $\alpha$)
are given by the following relations:
\begin{eqnarray}
n_s\sim1-\frac{2}{N_*}, \quad r\sim 12\frac{\alpha}{N_*^2}.
\end{eqnarray}
The dependency of $n_s$ and $r$ with respect to the number of $e$-foldings $N_*$ is quite common: it is shared by a lot of important models such as
the T-models themselves, the Starobinsky scenario 
(for $\alpha=1$) \cite{43,44}, the Goncharov-Linde model \cite{45}, the Higgs inflation \cite{46} or the so called string moduli inflation \cite{47,48,49,50}, because the shape of the potential during the inflationary stage provides $\epsilon_V\sim \eta_V^2$.
The plateau of the potential function is replaced by a pure quadratic shape if $\alpha>>160$.
However, we studied the local potential reconstruction (see \cite{51} for a review) of the $\textbf{E-models}$ with $n=1$ in \cite{37} which provides
\begin{eqnarray}
V(\phi)=\Lambda^4\left[1+d_1\left(\frac{\Delta\phi}{M_p}\right)+d_2\left(\frac{\Delta\phi^2}{M_p}\right)^2+...\right]
\end{eqnarray}
with
\begin{eqnarray}\label{eqn: estb}
\frac{d_2}{d_1}=-b, \quad \frac{\phi_*}{M_p}(b,d_1)=-\frac{1}{b} \ln \left( \frac{d_1}{2b} \right).
\end{eqnarray}
Starting from these definitions we can now reconstruct the postaccelerated epoch.

\section{Constraints on the reheating phase in early $\alpha$-attractor Universe }

\begin{figure}[htbp]
\centering
\includegraphics[width=8.6cm, height=6cm]{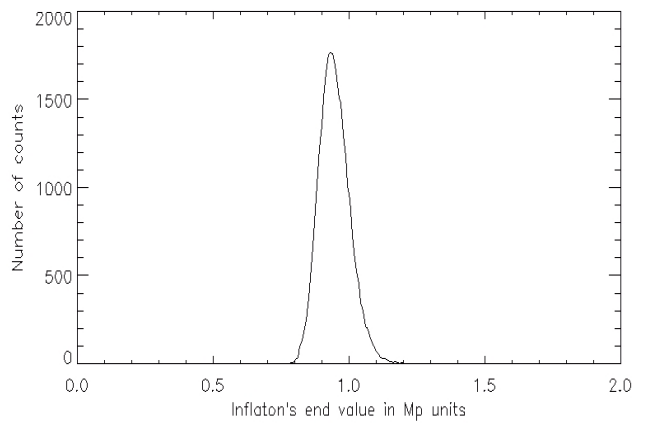}
\caption{Distribution of the variable $\phi_{end}$ when a mean value $\mu_{r}=0.003$ is detected.}
\label{fig: phi_end}
\end{figure}

The reconstruction of the postinflationary phase is possible when a robust statistical information is available on the input parameters
$\alpha$ and $\phi_*$.
For this goal, we simulate values of $n_s$ and $r$, randomly extracted from a multivariate Gaussian distribution of the form
\begin{equation}\label{eqn: gauss}
\mathcal{G}(n_s,r)=\frac{1}{\sqrt{4\pi^2\sigma^2_{n_s}\sigma^2_{r}(1-\rho^2)}} \exp{\left(-\frac{Q^2}{2}\right)}.
\end{equation}
Here
\begin{equation}
Q^2=\frac{1}{1-\rho^2}\hat{Q}^2 
\end{equation}
where
\begin{equation}
\hat{Q}=\left[\frac{(n_s-\mu_{n_s})^2}{\sigma_{n_s}^2} -2\rho\frac{(n_s-\mu_{n_s})(r-\mu_{r})}{\sigma_{n_s}\sigma_{r}}
+ \frac{(r-\mu_{r})^2}{\sigma_{r}^2} \right]
\end{equation}
The parameters $\mu_{n_s}$ and $\sigma_{n_s}$ are the mean and rms values of the scalar index while $\mu_{r}$ and $\sigma_{r}$ are the 
corresponding values for the tensor-to-scalar ratio.
The parameter $\rho$ is the correlation coefficient.
Let us start, considering $\mu_{n_s}=0.968$ and $\sigma_{n_s}=0.006$, compatibly with the current Planck bounds, 
and three different values for $r$ ($\mu_r=0.001,0.002,0.003$) with the same uncertainty $\sigma_r=0.0001$. 
We also assume $\rho=0.1$ to take into account our ignorance about possible correlations among $n_s$ and $r$.
The vacuum expectation value of the scalar field at the end of inflation is reconstructed through $\alpha$ (or $b$) by the Eq.($\ref{eqn: vev end}$).
In Tab.I we report the results for the three chosen values of $r$.
Note that, these results represent the extremal values that $\phi_{end}$ assumes in the considered cases: in principle inflation can stops before,
in general when $\epsilon_V < 1$ and this induces larger values for $\phi_{end}$.
In Fig.($\ref{fig: phi_end}$) we show the distribution of the values of $\phi_{end}$ associated with a detection $r=0.003$.
The variable $\phi_{end}$ is useful to derive the energy density of the Universe when inflation stops, $\rho_{end}$, by the relation Eq.($\ref{eqn: energydensity}$).
The number of $e$-foldings $N_*$ is reconstructed once $\phi_{end}$ and $\phi_*$ are known by Eq.($\ref{eqn: efolds complete}$).
Alternatively, one can also use Eq.($\ref{eqn: efolds app}$) if the condition $\phi_{end}\ll \phi_*$ is taken into account.
In Tab.II we report the related Monte Carlo results.
These kinds of $1$-$\sigma$ values on $N_*$ implies a great uncertainty
on the number of $e$-foldings $N_{reh}$ and especially, on the reheating temperature $T_{reh}$ due to Eq.($\ref{eqn: reh}$) and Eq.($\ref{eqn: gen_temp}$).


\begin{table}[htbp]
\begin{ruledtabular}
\begin{tabular}{lcdr}
\textrm{$\mu_{r}$}& \textrm{$\phi_{end}/M_p$ mean value}& \multicolumn{1}{c}\textrm{$\phi_{end}/M_p$ $1$-$\sigma$ value}\\
\colrule
0.001                     &    0.781                          &     0.061 \\                       
0.002                     &    0.889                          &     0.058  \\                       
0.003                     &    0.950                          &     0.056 \\                       
\end{tabular}
\end{ruledtabular}
\caption{Simulation results for the vacuum expectation value of the inflaton field when inflation stops. }
\label{tab: table1}
\end{table}


\begin{table}[htbp]
\begin{ruledtabular}
\begin{tabular}{lcdr}
\textrm{$\mu_{r}$}& \textrm{$N_*$ mean value}& \multicolumn{1}{c}\textrm{$N_*$ $1$-$\sigma$ value}\\
\colrule
0.001                       &    63.7                             &     13.5 \\                        
0.002                       &    63.0                             &     13.3 \\                       
0.003                       &    62.5                             &     13.1 \\                     
\end{tabular}
\end{ruledtabular}
\caption{Simulation results for the number of $e$-foldings $N_*$ before the end of inflation when $\sigma_{n_s}=0.006$.}
\label{tab: table2}
\end{table}


\begin{table*}
\begin{ruledtabular}
\begin{tabular}{ccccc}
                                   &\multicolumn{2}{c}{$w_{reh}=-1/3$}                          &\multicolumn{2}{c}{$w_{reh}=0$}              \\
    $\mu_{r}$           &          $N_{reh}$              &    $\ln T_{reh}/M_p$          &       $N_{reh}$           &       $\ln T_{reh}/M_p$     \\ \hline
 $0.001$          &          $3.70\pm1.95$          &       $-8.89\pm0.97$          &       $7.40\pm3.90$       &      $-12.59\pm 2.91$       \\
 $0.002$          &          $5.27\pm1.90$          &       $-9.57\pm0.94$          &       $10.55\pm3.81$      &      $-14.84\pm2.85$         \\
 $0.003$          &          $6.46\pm1.87$          &       $-10.11\pm0.93$         &       $12.93\pm3.75$      &      $-16.57\pm2.81$         \\
 $0.010$          &          $12.02\pm1.73$         &       $-12.76\pm0.86$         &       $24.02\pm3.47$      &      $-24.77\pm2.59$         \\
 $0.013$          &          $13.95\pm1.67$         &       $-13.72\pm0.83$         &       $27.89\pm3.35$      &      $-27.67\pm2.51$          \\
 $0.016$          &          $15.86\pm1.61$         &       $-14.67\pm0.78$         &       $31.73\pm3.22$      &      $-30.54\pm2.41$          \\
\end{tabular} 
\end{ruledtabular}
\caption{Simulation results for the duration of the reheating phase and the reheating temperature related to a detection of the scalar spectral index $n_s=0.9650$
for detection of the tensor-to-scalar ratio of different orders.
The sign of the function $f$ is positive so are allowed mean value of the EoS $w_{reh}<1/3$.
We report the results for two different mean value of EoS: $w_{reh}=-1/3$ and $w_{reh}=0$.}
\label{tab: table3}
\end{table*}

\begin{table*}
\begin{ruledtabular}
\begin{tabular}{ccccc}
                                   &\multicolumn{2}{c}{$w_{reh}=2/3$}                          &\multicolumn{2}{c}{$w_{reh}=1$}              \\
    $\mu_{r}$           &          $N_{reh}$              &    $\ln T_{reh}/M_p$          &       $N_{reh}$           &       $\ln T_{reh}/M_p$     \\ \hline
 $0.001$          &          $13.43\pm4.65$         &     $-23.84\pm5.81$           &       $6.71\pm2.32$       &      $-17.13\pm 3.49$       \\
 $0.002$          &          $9.95\pm4.55$          &     $-19.37\pm 5.70$          &       $4.97\pm2.27$       &      $-14.40\pm3.42$         \\
 $0.003$          &          $7.27\pm4.47$          &     $-15.98\pm5.60$           &       $3.63\pm2.24$       &      $-12.35\pm3.36$         \\
\end{tabular} 
\end{ruledtabular}
\caption{Simulation results for the duration of the reheating phase and the reheating temperature related to a detection of the scalar spectral index $n_s=0.9680$
and a tensor-to-scalar ratio $r\sim 10^{-3}$.
The sign of the function $f$ is negative so are allowed large value of the EoS $w_{reh}>1/3$.
We report the results for: $w_{reh}=2/3$ and $w_{reh}=1$.}
\label{tab: table4}
\end{table*}

\begin{table*}
\begin{ruledtabular}
\begin{tabular}{ccccc}
                                   &\multicolumn{2}{c}{$w_{reh}=-1/3$}                          &\multicolumn{2}{c}{$w_{reh}=0$}              \\
    $\mu_{r}$           &          $N_{reh}$             &    $\ln T_{reh}/M_p$         &       $N_{reh}$           &       $\ln T_{reh}/M_p$     \\ \hline
 $0.010$          &          $2.68\pm2.05$         &       $-8.13\pm1.02$         &       $5.36\pm4.10$       &      $-10.81\pm3.01$         \\
 $0.013$          &          $4.96\pm1.97$         &       $-9.23\pm0.98$         &       $9.92\pm3.95$       &      $-14.21\pm2.95$          \\
 $0.016$          &          $7.26\pm1.88$         &       $-10.40\pm0.93$        &       $14.52\pm3.76$      &      $-17.67\pm2.82$          \\
\end{tabular} 
\end{ruledtabular}
\caption{Simulation results for the duration of the reheating phase and the reheating temperature related to a detection of the scalar spectral index $n_s=0.9680$
for a detection of the variable $r$ of the order of $10^{-2}$.
In this case, the sign of the function $f$ is still positive, therefore $w_{reh}<1/3$ are allowed.
We report the results: $w_{reh}=-1/3$ and $w_{reh}=0$.}
\label{tab: table5}
\end{table*}

This piece of evidence suggests that improvements of the estimation on $n_s$ could be strongly reduce the uncertainty on the inflationary number $e$-folds $N_*$.
Nowadays, foreseen cosmological experiments aim to have an uncertainty $\sigma_{n_s}\sim 0.002$ or even better.
In this paper we try to be optimistic and we set $\sigma_{n_s}=0.0006$.
As we will discuss in the last section, it is hard to get such a sensitivity from a single experiment but rather by combining future several experiments.
We assume two possible scenarios for the scalar spectral index $n_s$
\begin{itemize}
\item $\mu_{n_s}=0.9650$, $\sigma_{n_s}=0.0006$
\item $\mu_{n_s}=0.9680$, $\sigma_{n_s}=0.0006$
\end{itemize}
and for both cases we consider two possible realizations for the tensor-to-scalar-ratio
\begin{itemize}
\item $\mu_r=0.001,0.002,0.003$
\item $\mu_r=0.010,0.013,0.016$
\end{itemize}
with $\sigma_r=0.0001$ and $\rho=0.1$.
In the analysis we stress what kinds of EoSp are allowed for each cases ($w_{reh}$ larger or less then $1/3$).
As we can see from Tab.III, in the first case $n_s=0.9650$ there is an independence on the order of magnitude that the mean value of ${r}$ takes on.
In all considered cases of $r$, the function $f$ results to be positive and low values of the EoS, $w_{reh}<1/3$, are allowed.
Once the parameter called $\mu_{r}$ results to be setted then, the mean value of $N_{reh}$ increases as $w_{reh}$ gets larger.
Consequently, the final reheating temperature gets lower.
When we increase the value of the scalar spectral index up to $n_s=0.9680$, we deal with two different situations.
In the range of $r\sim 10^{-3}$, we have $f<0$ so the reheating is characterized
by a large mean value of the EoS, $w_{reh}>1/3$ (Tab.IV).
In this case, once $\mu_r$ results are fixed, $N_{reh}$ gets lower as $w_{reh}$ increases.
Therefore, the reheating temperature becomes larger.
In the range of $r\sim 10^{-2}$, instead, we have $f>0$ and so the required
values of the EoS turns to be $w_{reh}<1/3$.
Consequently, we recover the same behavior seen when $n_s=0.9650$ (see Tab.V).

\section{Discussion and conclusions}

In this paper, we reconstructed some aspects of the postinflationary phase in the context of superconformal $\alpha$-attractor (supergravity) models of inflation.
In particular, we focused our study on constraining the duration of the reheating era, $N_{reh}$, and the re-heat scale of the Universe,
i.e. the reheating temperature, $T_{reh}$.
The reconstruction of these two parameters is strongly related to the mean value of the EoSp of the reheating fluid, $w_{reh}$.
In our study we have outlined the existence of transitions.
As we pass from a mean value of the scalar index $n_s=0.9650$ to $n_s=0.9680$ in the range of $r\sim 10^{-3}$, then we have a change
in the sign of the function $f$: from $f>0$ to $f<0$.
This corresponds to a change of the allowed EoS from $w_{reh}<1/3$ to $w_{reh}>1/3$.
A this point, moving toward larger values of $r$ the function $f$ turns to be progressively positive.
The case of $w_{reh}>1/3$ is not favored from the standpoint of QFT but it still remain a possibility and in this sense it represents a compelling challenge
in the context of early fundamental physics, as outlined also in \cite{40}.
Scenarios in which $w_{reh}<1/3$ are characterized by a specific property.
Once $\mu_{r}$ is fixed, the duration of the reheating gets larger as $w_{reh}$ increases and the reheating temperature gets lower.
The same happen if we fix $w_{reh}$ with a varying $\mu_{r}$.
In the case in which $w_{reh}>1/3$ we have the opposite behavior as summarized in the previous section.
Anyway, we want to stress that these results are overall in agreement with the analytic studies done by Eshaghi et all as well as Ueno and Yamamoto \cite{42}.
For example, a detecion of $n_s=0.965$ and $r\sim 0.016$ ($\alpha\leq 5$) with a standard equation of state $w_{reh}=0$, implies
a very low reheating temperature of the order of $T_{reh} \sim 6\times 10^{8}$ GeV.
At the same time a detection of $n_s=0.968$ and $r\sim 0.016$ again with a standard equation of state, implies a reheating temperature
$T_{reh} \sim 10^{10}$ GeV.
Therefore we need $\alpha>1$, in order to move toward prolonged reheating 
with low temperatures, tipically $\alpha>5$ as outlined especially by Eshaghi et all \cite{42}. 
Low values of the tensor to scalar ratio with $n_s=0.9680$ are not consistent with standard equation of state and show 
an opposite behaviour:
we have low reheating temperature for low values of $\alpha$.
Another important aspects of this work is in the resulting distributions for $N_{reh}$.
Providing robust constraints on the duration of the reheating is fundamental for understanding how much important is the hypothetical preheating phase.
To have an efficient preheating phase we need $N_ {reh}\sim 10$ or more.
In some cases, (see Tab.V for $w_{reh}=-1/3$, for instance) the $N_{reh}$ is tiny, so the possibility for a preheating seems to be not so significantly
even considering the relative errors.
On the other hand, the evidence of a prolonged reheating phases implies very low reheating temperatures.
As well summarized by Dai, Kamionkowsi and Wang in \cite{40}, there could be tensor fluctuations produced during inflation
which reenter in the microphysical or Hubble horizon during the cosmic reheating history.
These fluctuations therefore were stretched on very small scales compared to the current Hubble radius $c/H_0$.
In particular, these GW should be locally detectable on scales of the order of the solar system, about $40$-$e$folds below the CMB scales
as suggested in \cite{52,53}, for instance.
Therefore, they could bring fundamental cosmological and particle physics information about the reheating phase.
The robustness of the presented analysis is related to the power of forthcoming future experiments to improve the current datasets.
The first important point is in the capability of efficiently exploring ranges of the tensor-to-scalar ratio below the PLANCK mission sensitivity.
The second important point is in the capability of reducing the uncertainty on the scalar spectral index $n_s$.
As we saw in the previous section, our forecasting are quite sensitive to some optimal constraint on $n_s$.
In fact, in our simulations we assume that we will be able to constrain $n_s$ up to the order of $6\times 10^{-4}$ whereas the actual limit given by PLANCK are 
of the order of $6\times 10^{-3}$.
This improvement, looking at the next-generation experiments could be a reachable goal.
Indeed, missions such as EPIC \cite{4} or CoRE (LiteCore) \cite{5} combined with results coming from experiments like PRISM \cite{8}, EUCLID \cite{54} or by the 21-cm surveys \cite{55,56} could able to reduce the total uncertainty on $n_s$ at the required level of our analysis.

\begin{acknowledgments}
This work has been supported in part by the STaI ``Uncovering Excellence" grant of the University of Rome
``Tor Vergata", CUP E82L15000300005.
A.D.M would like to thank Nicola Bartolo and Sabino Matarrese for the useful discussion on the early version of the paper.
P.C thanks ``L'isola che non c'\`e S.r.l" for the support.
\end{acknowledgments}




\nocite{*}

\bibliography{apssamp}

\end{document}